# Planar Multilayered 2D GeAs Schottky Photodiode for High Performance VIS-NIR Photodetection

Ghada Dushaq and Mahmoud Rasras

(Department of Electrical and Computer Engineering, New York University Abu Dhabi, UAE)

*ghd1@nyu.edu , *mr5098@nyu.edu

Novel group IV–V 2D semiconductors (e.g., GeAs and SiAs) has arisen as an attractive candidate for broadband photodetection and optoelectronic applications. This 2D family has wide tunable bandgap, excellent thermodynamic stability, and strong in-plane anisotropy. However, their photonic and optoelectronic properties have not been extensively explored so far. In this work we demonstrate a broadband back-to-back metal-semiconductor-metal (MSM) Schottky photodiode with asymmetric contact geometries based on multilayered 2D GeAs. The photodetector exhibited a Schottky barrier height (SBH) in the range of 0.40 to 0.49 eV. Additionally, it showed low dark current of 1.8 nA with stable, reproducible, and excellent broadband spectral response from UV to optical communication wavelengths. The highest measured responsivity in the visible is 905 A/W at 660 nm wavelength and 98 A/W for 1064 nm near infrared. Most notably, the planner configuration of this GeAs photodetector showed low detector capacitance below 1.2 pf, low voltage operation (<1V), and a large bandwidth which may exceed 40 GHz. The stability and broadband response of the device are promising for this 2D material's application in high-speed optoelectronic devices.

## 1. Introduction

Narrow bandgap semiconductors play a major role as building blocks in providing solutions for emerging challenges in optoelectronics [1–4]. For instance, light detection in the near-infrared (NIR) region is of particular scientific interest and technological importance [5]. They open new venues for imperative applications, such as molecular imaging, remote sensing, free space telecommunication, and optical radar. State-of–the-art infrared and broadband detectors are mostly based on semiconductors with narrow band gaps, such as HgCdTe alloy [6], InSb [7], or quantum-well and quantum-dot structures on group III to V materials [8,9]. However, these materials suffer from major challenges such as toxicity, growth difficulties (lattice mismatch) [10,11], complex advanced fabrication processes requirements, and a low operating temperature that limits their wide applications. Therefore, exploring new materials with excellent optoelectronic properties for multiband response ranging from ultraviolet to near-infrared is highly desirable.

Recently, two dimensional materials, such as graphene, transition metal dichalcogenides, and black phosphorus (BP), have exhibited the most potential as semiconductor materials candidates for future optoelectronic applications due to their extraordinary properties [12,13]. Owing to their layer-dependent and appropriately sized bandgaps, photodetectors based on various 2D materials have been successfully demonstrated [14–17]. Additionally, the dangling-bond-free surface and weak van der Waals (vdW) interaction provide excellent electronic properties, and allow substrate integration without the traditional constraints of lattice mismatch and material incompatibilities. Graphene-based optoelectronics have great potential due to its high mobility and excellent mechanical properties [18,19]. However, achieving high responsivity and ultrafast response is still challenging due to its zero bandgap [20,21]. Other 2D materials, such as black phosphorus (BP) exhibits a unique electronic structure with attractive broadband absorption properties from visible to mid-infrared regions and ultrashort recovery time of several tens of femtoseconds [22,23]. However, BP is unstable in air and degrades easily to phosphoric acid under the action of light, oxygen, and water

[24,25]. Therefore, the development of novel materials with tunable band gap, broadband absorption, strong optical response, and excellent air-stability is still a long-term goal.

Very recently, novel group IV–V 2D compounds (e.g., GeAs and SiAs) have arisen as an attractive candidate for broad-band photodetection and optoelectronic applications [26,27]. This 2D family are closely related to traditional group IV (e.g., Si, Ge) and group III–V semiconductors (e.g., GaAs). They exhibit tunable bandgaps (Eg), excellent chemical stability, and strong in-plane anisotropy [28–30]. Despite their superior photonic and optoelectronic properties, much of the research are focused on theoretical studies [31–33], whereas experimental exploration is extremely rare [34,35].

Herein, we report high-performance and broadband back-to-back metal-semiconductor-metal photodetector (MSM) Schottky photodiode based on multilayer 2D GeAs on $SiO_2$/Si substrates with asymmetric contact geometries. The room temperature photoluminescent measurements showed a thickness dependent bandgap ranging from 0.77 eV to 2.2 eV. Additionally, results revealed that Au/Cr–GeAs interface is dominated by a large Schottky barrier height (SBH) of ~ 0.4-0.49 eV and a current on/off ratio (rectification ratio) of $10^4$. The photodiodes exhibited low dark currents and excellent broadband spectral responses from ultraviolet to optical communication wavelengths with excellent stability and reproducibility.

## 2. Experimental results

### 2.1 Structural and Optical Properties of Multilayered 2D GeAs

Bulk GeAs belongs to the space group C2/m, where, it crystallizes in layered GaTe structure type [31,34]. Figure 1(a) shows a schematic of a 2D GeAs crystal structure. Within each layer, each Ge atom (red) bonds with one Ge atom and three As atoms (black), forming distorted octahedra. Furthermore, the GeAs layer interacts with each other by weak van der Waals forces and all layers are terminated by As atoms. The crystal structure exhibits a high anisotropic nature due to its two distinguishable types of Ge-Ge bonds, one type is parallel and the other is perpendicular to the layer plane. In order to identify the layered structure of the GeAs crystal, the GeAs flakes were exfoliated from their natural crystals and deposited onto a $SiO_2$/Si substrate. The flakes are identified under an optical microscope and their topography is characterized by atomic force microscope (AFM). An optical image shown in Fig. 1 (b) reveals a distinct contrast of a multilayered GeAs flake that is successfully exfoliated on SiO$_2$ substrate. Figure 1(c) depicts its height profile obtained from an AFM image scan. We chose GeAs flakes thickness ranging from 32 to 60 nm for the device fabrication which offers a trade-off between high photo-absorption and low dark current (Figure S1 shows extra optical image of multilayered 2D GeAs and AFM scan of exfoliated flakes).

Confocal Raman spectroscopy enables direct identification of material's vibrational modes and a powerful tool for determining the crystal symmetries. The Raman and photoluminescence (PL) measurements were performed in ambient air environment with excitation laser lines of 488 nm and 532 nm, respectively. The optical power of the excitation laser lines was kept below 1 mW to avoid damage to the GeAs. The Raman scattering was collected by an Olympus 100x objective and dispersed by 1,800 and 600 lines per mm gratings for the Raman and PL measurements, respectively. Figure 1(d) illustrates the measured spectra of a multilayer GeAs flakes under un-polarized 488 nm excitation source. The Raman spectrum reveals several peaks attributed to the highly asymmetric crystal structure. Furthermore, two intense peaks at 169 cm$^{-1}$ and 255 cm$^{-1}$ related to the atomic vibration perpendicular to the b-axis were observed. This is consistent with previous observation of the Raman spectra of 2D GeAs [35]. The Raman peaks at 142, 169.3, 269, 298 cm$^{-1}$ belong to A$_g$ modes, and those at 243.5, 254.8 cm$^{-1}$ belong to B$_g$ modes.

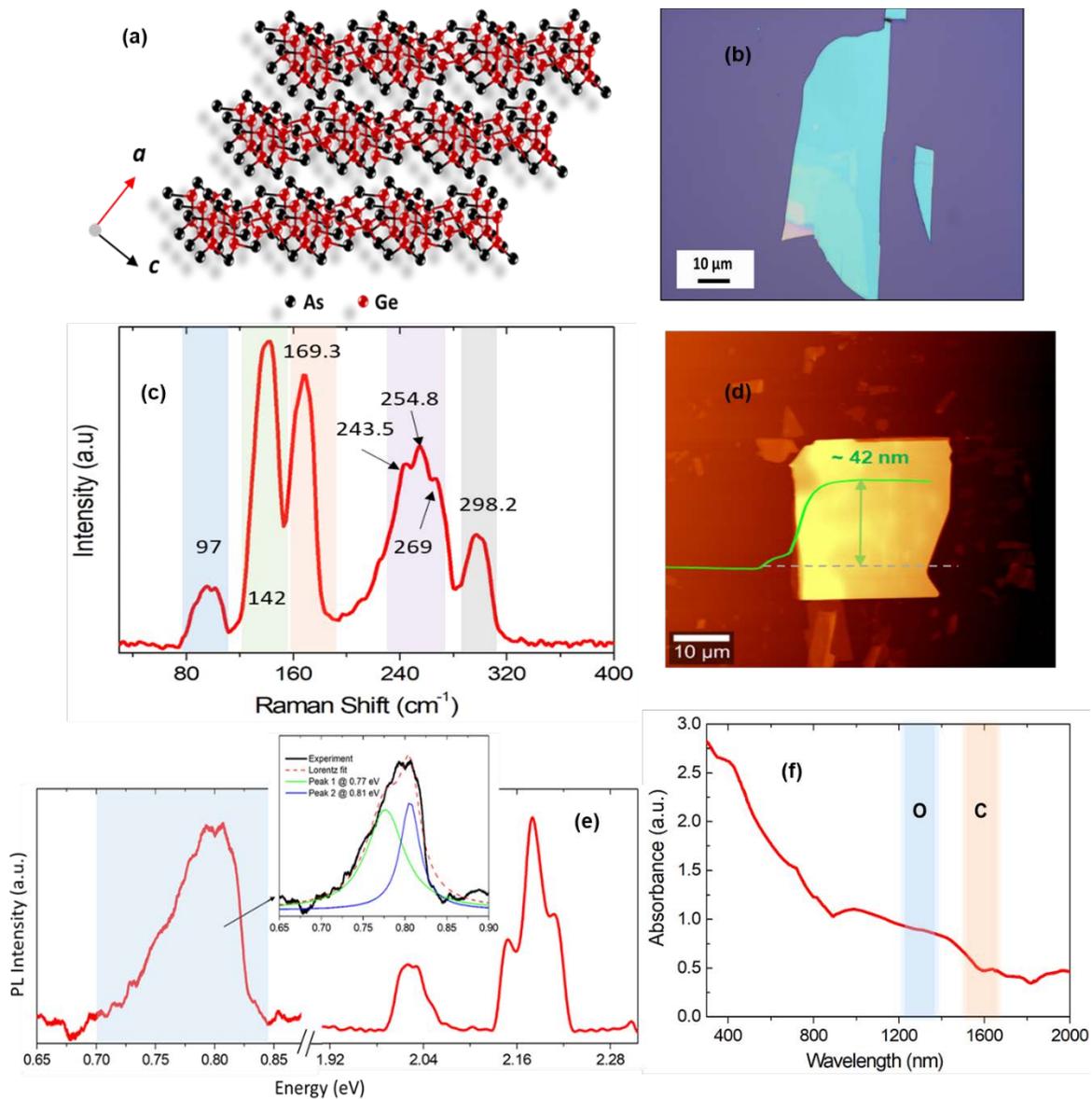

Fig. 1. 2D GeAs crystal structure, schematic and material characterization (a) schematic of 2D GeAs crystal structure (b) optical image of multilayered GeAs on 300nm $SiO_2$/Si substrate (c) Raman spectrum of multilayer GeAs flake using un-polarized 488 nm laser excitation (d) AFM image of GeAs flake showing 42 nm flake thickness (e) photoluminescence spectra (PL) of multilayered GeAs using 532 nm excitation (f) absorption spectra of 2D GeAs (O band and C band).

In order to gain insight into the nature of the optical transitions and the material bandgap, we measured the photoluminescence (PL) spectra for different GeAs crystal thicknesses. Figure 1(e) shows the PL spectra of a 42 nm thick flake exfoliated on a $SiO_2$/Si substrate. Multiple emission peaks in the visible (2.0-2.2 eV) and a single emission peak in the near infrared (NIR), were observed with a minimum at a photon energy of 0.77 eV and a FWHM of 53 meV. Figure S2 shows the optical images of the flakes measured in the PL experiment, the spectra showed no NIR peak for flakes with thickness < 10nm while NIR peak are measured for thickness in the range of 32nm-82nm. The NIR PL were identified in two superimposed peaks using multi-peak Lorentz- Gaussians fitting. The energy shift between the two peaks is about 40 meV. We do not observe any additional emission peaks in the PL spectra of thicker layers up to 82 nm. The measured bandgap values are close to the theoretical predicted results [28]. To further confirm the narrow band gap nature of multilayer GeAs, we measure its UV-VIS-NIR absorption spectra as shown in Fig.1 (f). A drop of absorption (Transmission of ~ 31%) is observed starting from wavelength of 1.6 μm, which confirms mid-infrared absorption resulting from its small band gap.

*2.2 Device fabrication*

A 300 nm $SiO_2$/ Si substrates were precleaned using acetone and isopropyl alcohol at room temperature. This was followed by a rinse and drain in de-ionized water, and the substrates were subsequently dried by using $N_2$. Multilayer GeAs flakes were exfoliated from their natural crystals (commercially available from 2D Semiconductor) and deposited on the substrates. Following the GeAs flakes exfoliation, metal electrodes were defined using standard photolithography and lift-off process. Metal contacts with different illumination areas (100 –600 μm$^2$ ) were fabricated by depositing a 25 nm Cr/100 nm Au film by a thermal evaporator using high purity Au and Cr pellets.

*2.3 MSM photodetector (MSMPD) operation principle*

A Metal-semiconductor-metal photodetector (MSMPD) structure has a planar design which offers low capacitance suitable for high-speed detection. Additionally, the simplicity of the fabrication process and its low-voltage operation make it ideal platform for device integration. The vertically illuminated detector consists of two metal-semiconductor contacts connected back-to-back or back-to-back diodes as shown in Fig. 2(a). The band diagram of the MSM detector under thermal equilibrium is shown in Fig. 2(b). The electron barrier height at each metal side is $\phi_1$ and $\phi_2$, respectively. If the metal contacts area are similar, then $\phi_1= \phi_2$, this case is called symmetric MSM detector. When a bias is applied to the MSM structure (Fig. 2(c)) in such a way that one contact is reverse biased and the other is grounded, the applied bias increases the potential barrier height for the electrons and the depletion region width increases. The incident photons are detected by generating photocarriers which drift under the influence of the applied electric field between the contacts. The details of the metal/GeAs Schottky barrier height (SBH) and electrical measurements will be discussed in the next section.

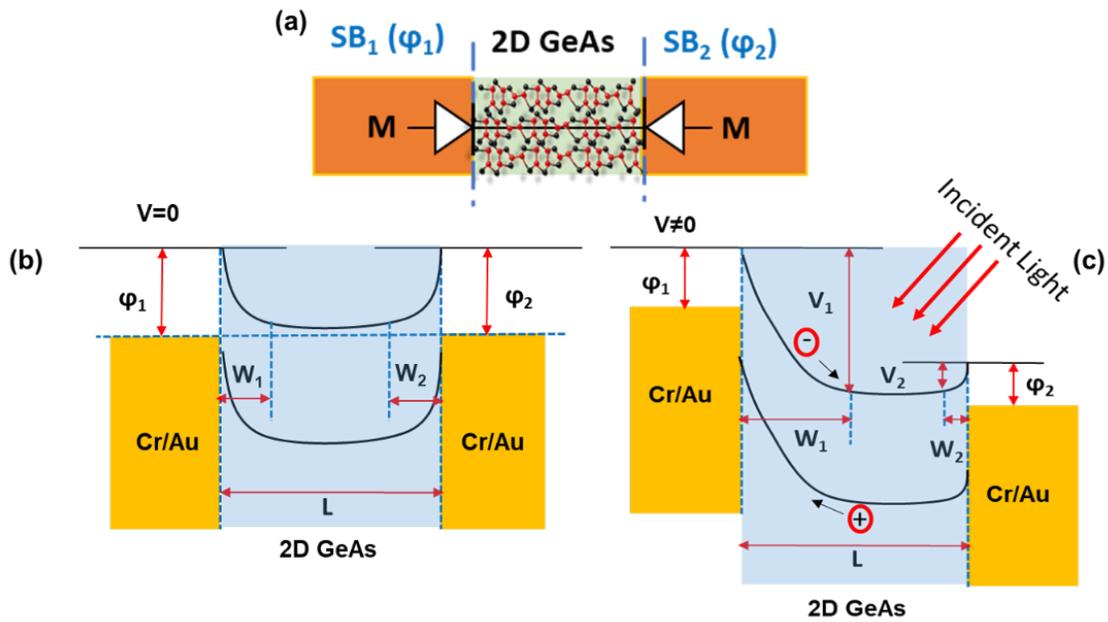

Fig. 2. 2D GeAs metal-semiconductor-metal photodetector configuration (a) schematic of back-to-back diode where Schottky diode is formed at the metal/2D GeAs junction (b) band diagram of MSM PD at zero voltage (c) under electrical bias.

3. **Device characterization**

*3.1 Electrical measurements*

The dark current-voltage (IV) characteristic of GeAs Schottky photodiode with asymmetric contact geometries and at zero back-gate voltage is shown in Fig. 3(a). In the MSM structure, different contact lengths are chosen for the two metal–semiconductor electrodes, thus forming an asymmetric contact geometry (see Figure S3). The metal/GeAs junction exhibited a typical rectifying diode behavior with a leakage current in the order of 1.8 - 6 nA at a reverse bias of 1V. The high energy barriers naturally formed at the interfaces of the metal/semiconductor junction can effectively reduce the dark current noise. Additionally, a rectification ratio of ~$10^4$ under ±3 V operation condition was obtained with electrode length difference (ΔL) of ~ 41um. It is important to point out that the rectification ratio induced by asymmetrical contacts depends on two important factors, the degree of contact asymmetry (see Figure S3) and the metal-semiconductor barrier height [36]. The above electrical parameters are very comparable to those of 2D van der Waals heterojunction diodes previously reported in literature [37–39]. Additionally, with back-gate device geometry, p-type behavior are observed at room temperature. This is in agreement with previously published reports[34].

In order to gain more insights into the intrinsic electronic transport properties of GeAs and the metal-semiconductor contact barrier height, temperature-dependent electrical measurements were carried out as shown in Fig. 3(b). These characteristics generally dominate the electronic transport in Schottky photodiode types. The temperature-dependent current of a Schottky junction is described by the thermionic emission theory of majority carrier over the Schottky barrier according to [40,41]:

$$J(T) = J_s(T)\left[exp\left(\frac{qv}{nk_BT}\right) - 1\right] \qquad [1]$$

Where $n$ is the ideality factor, $k$ is Boltzmann's constant ($k = 8.62 \times 10^{-5}$ eV/K), $T$ is the temperature in Kelvin, $q$ is the electronic charge ($1.6 \times 10^{-19}$ C), and the saturation current density $J_s(T)$ is given by [40]:

$$J_S(T) = A^*T^2 exp\left(-\frac{q\,(SBH)}{k_BT}\right) \qquad [2]$$

Where $A^*$ is Richardson constant.

We extracted the effective SBH from the temperature-dependent IV measurements described in the Richardson plot of $\ln(I/T^2)$ versus $1000/T$ (see Fig. 3(c)) [42] and from the slope at the forward bias linear region of $\ln(J) - V$ curve [43]. The estimated SBH of Au/Cr/2D GeAs photodiode is in the range of 0.40 to 0.49 eV.

The photodetector bandwidth is a key figure of merit in high-speed optical applications [44,45]. Typically, the resistor-capacitor (RC) constant limits the operation speed of a photodetector in a practical circuit, where the 3-dB electrical bandwidth (BW) is related to the RC constant by $1/2\pi RC$ [46]. Figure 3(d) shows the capacitance-voltage and bandwidth-voltage characteristics of the 2D GeAs MSMPD. A decrease in the junction capacitance with the reverse bias is observed. We measured device capacitance values in the range of 0.1–1.2 pF for applied voltages of −1 to 1 V, respectively. By considering a series resistance of ≈ 50 Ω, a relatively high bandwidth of ≈ 40 GHz can be achieved. In particular, it is worth noting that 2D IV–V materials feature high mobility for both free electrons and holes [33]. These results collectively demonstrate the excellent characteristics of GeAs MSMPD.

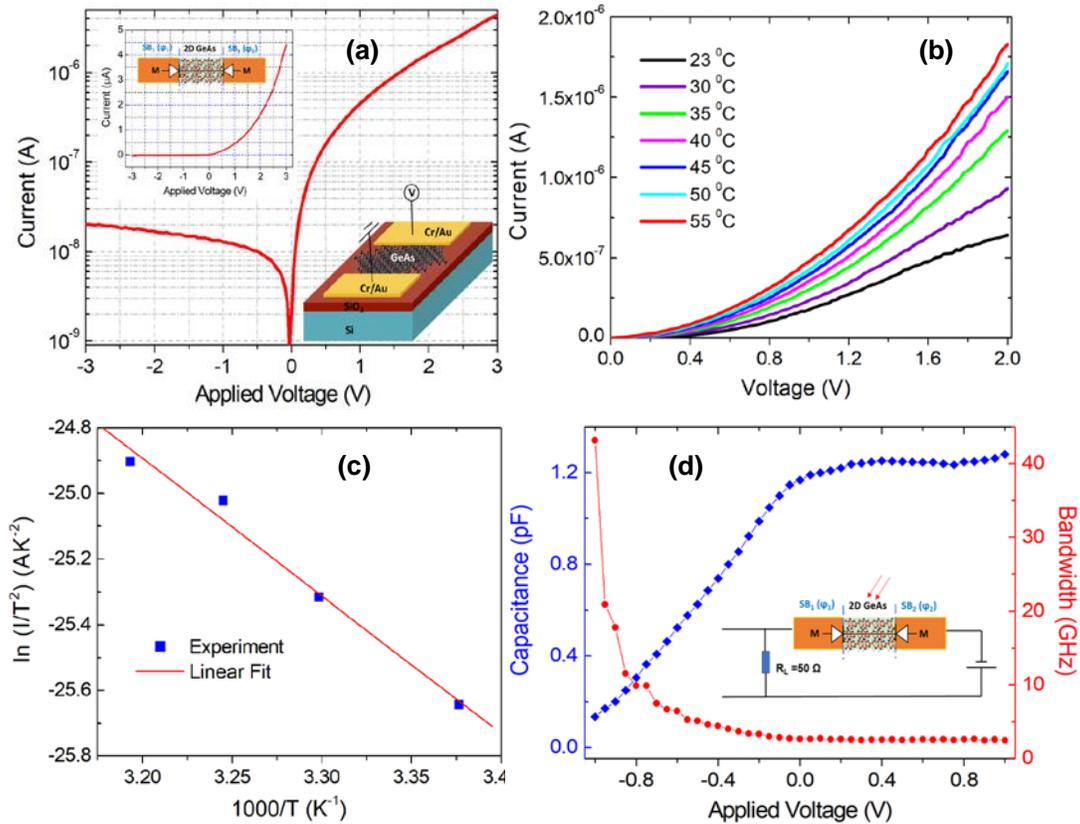

Fig. 3. Electrical characteristic of 2D GeAs Schottky photodiode (a) dark IV characteristic on a logarithmic scale showing diode rectification behavior, insets showing the linear scale IV and schematic of the 2D GeAs MSMPDs (b) temperature dependent IV characteristic from room temperature to 55º C (c) Richardson plot of ln (I/T²) versus 1000/T for Cr/Au and GeAs contact (d) Capacitance and bandwidth versus reverse bias characteristics of 2D GeAs MSMPDs.

*3.2 Optical Measurements*

To investigate the photoresponse properties of GeAs MSMPDs, the devices were illuminated by different laser diodes (LDs) with wavelengths of 406, 532, 660, 1064, and 1310 nm. Optical power densities of all LDs were calibrated before the measurements using standard photodiode power sensor. Figures 4(a) and 4(b) display the dark and photo-induced I–V characteristics of the GeAs Schottky diode on a logarithmic scale (Linear scale is shown in Figure S4). As can be seen, the current at reverse bias of the photodiode increased sharply under illumination, leading to high $I_{on}/I_{off}$ ratios. It is worth noting that the GeAs photodiode also exhibited photovoltaic (PV) behaviors (see Fig. 4 (a) and (b)). An open circuit voltage ($V_{oc}$) of 0.2–0.3V and a short circuit current ($J_{sc}$) of 10–40 nA were observed. This indicates that the device has also a potential for use in solar cells or photovoltaic type detectors [47]. The self-driven property and its dependence on the degree of contact asymmetry has been previously demonstrated on 2D $WSe_2$ system.[36]

The spectral responsivity as a function of the illumination wavelengths is plotted in Fig. 4(c). The GeAs MSMPDs exhibited high photosensitivity over a broad wavelength range from UV (406 nm) to near-infrared NIR (1310 nm) with stable and repeatable response. This observation is in accordance with the absorption spectra described in Fig. 1(f) and Fig. 4(c), which indicates that the device can be operated in the near-infrared regime. Additionally, we investigated the stability and the photoresponse of the same photodetectors after 14 days, they demonstrated the same performance without photocurrent degradation. A peak responsivity of 905.5 A/W was observed under red laser diode (LD) illumination (660 nm). Although the responsivity decreases slowly with increasing wavelength, the device exhibited a relatively high responsivity of 21.3 A/W at 1310 nm, indicating its potential for applications in the fiber optics O-band communication. It is worth noting that the relatively large flake thickness plays a beneficial role in improving the optical absorption and thus the responsivity of the GeAs MSMPDs. It is expected that the responsivity can be further enhanced by optimizing the device configuration by integrating it in a waveguide-based photodetector [17].

Figure 4(d) shows the measured photocurrent versus applied reverse bias under 1064 nm and 1310 nm LD wavelengths illumination. As can be seen, the current at zero voltage greatly increased from 1.8 nA (dark current) to 2.5 µA at 1064 nm, and 1.4 µA at 1310 nm, exhibiting an $I_{on}/I_{off}$ ratios of $1.4 \times 10^3$ and $7.8 \times 10^2$, respectively. Therefore, we can deduce that the GeAs MSMPDs can function as a self-powered photodetector.

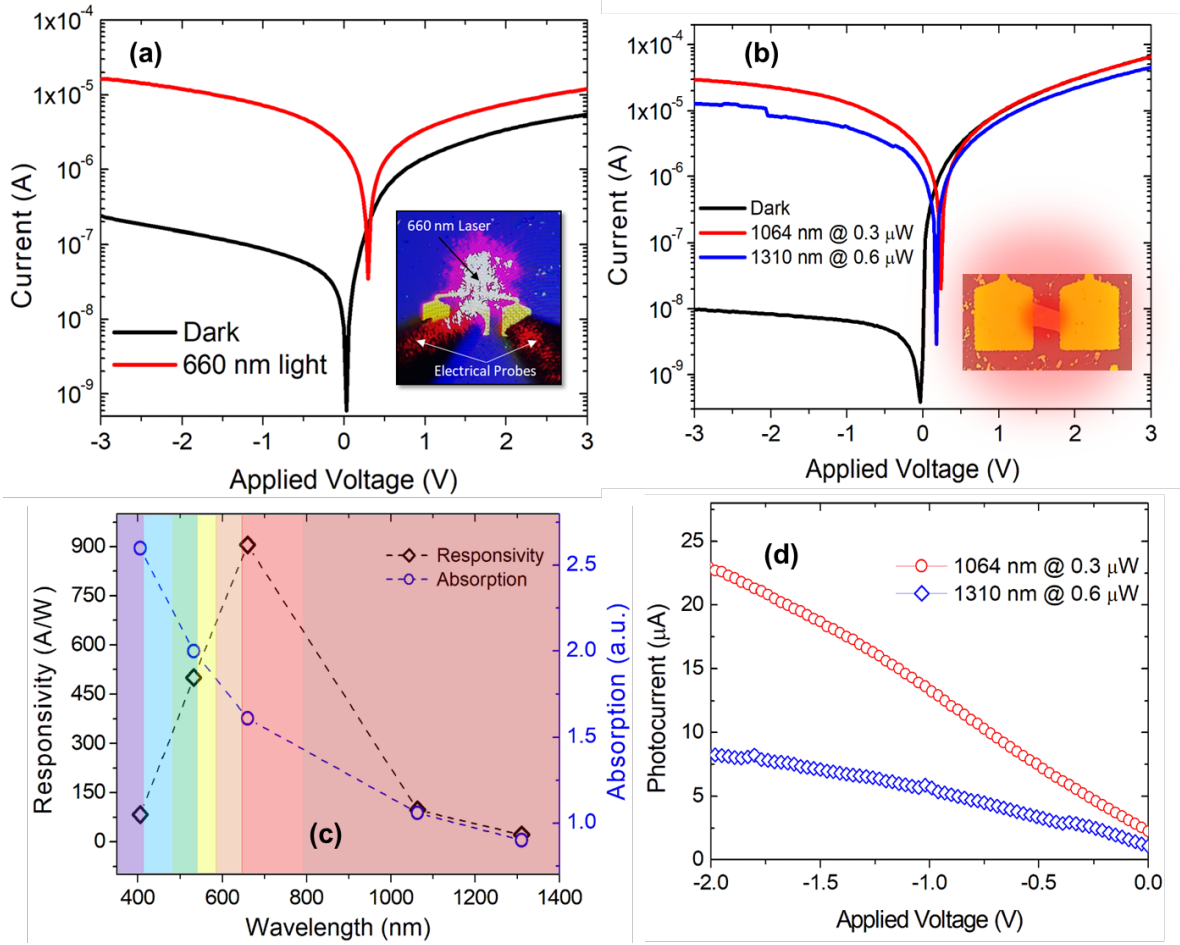

Fig. 4. Photo-induced characteristic of GeAs Schottky photodiode (a) I-V curves on a logarithmic scale under 660 nm laser illumination along with their dark state, insets showing the device under test (b) I-V curves on a logarithmic scale under 1064 nm and 1310 nm lasers illumination along with their dark state, inset showing the device structure under test (c) wavelength-dependent responsivities (d) photocurrent versus reverse bias voltage at 1064 nm and 1310 nm.

The specific detectivity ($D^*$) parameter determines the minimum illuminating light power that a detector can distinguish from the noise [48]. Figure 5(a) depicts the detectivity versus applied voltage measured at 660 nm LD illumination. A value of 8.6 x$10^{12}$ Jone at -3V is obtained. To compare our 2D GeAs photodetector performance to other reported 2D semiconductors [49–51], Fig. 5(b) depicts their spectral response. It clearly shows that our device exhibits a broadband photo-response comparable to that reported of BP [52] and it has a broader response compared to TMDC-based photodetector. This spectral response of 2D GeAs is very promising for application in future NIR-MIR optoelectronic devices.

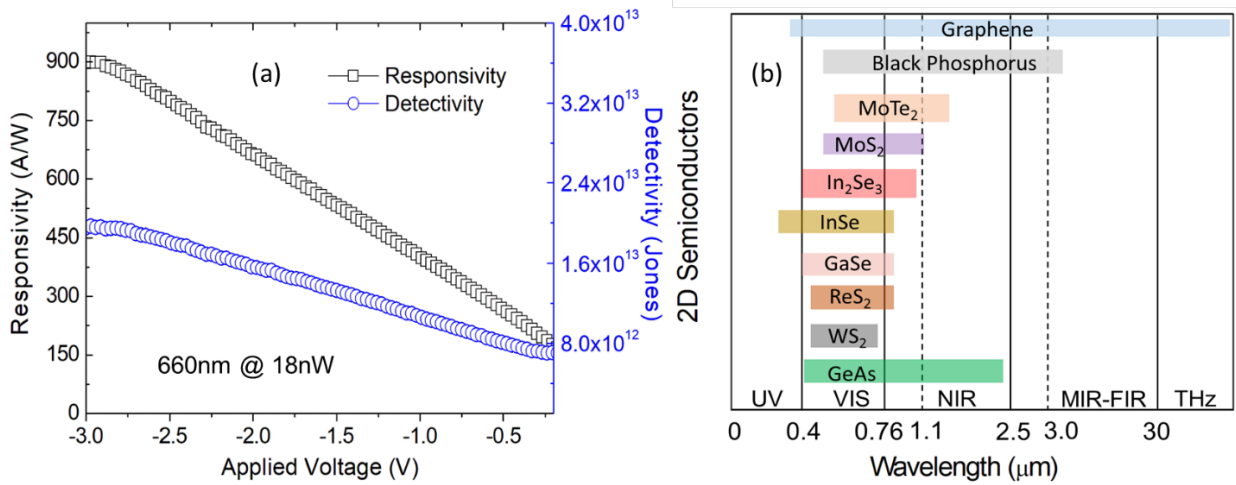

Fig.5 (a) Responsivity and detectivity versus reverse bias voltage at 660 nm (b) spectral response range for some of 2D semiconductors.

4. Conclusion

In summary, we have demonstrated a broadband back-to-back Schottky photodiode based on multilayered 2D GeAs. The photodiode showed an excellent performance and a high rectifying (On/Off) ratio over $10^4$. Additionally, a SBH of 0.40 to 0.49 eV was formed between the Au/Cr/GeAs junction which has effectively reduced the detectors dark noise. Remarkably, the as-assembled device exhibited a broadband photosensitivity from the UV to the optical communication wavelengths (NIR, 1310 nm) with the responsivity reaching 21.3 A/W at 1310 nm. In addition, a stable and repeatable photo response is observed. Notably, an obvious photovoltaic behaviors under illumination were also detected. The planner configuration of the devices offers low detector capacitance, low voltage operation and large bandwidth which may exceed 40 GHz. This enables fast separation and transportation of photogenerated carriers. In conclusion, 2D GeAs Schottky photodiodes can be promising building blocks for advanced high-speed optoelectronic applications.

**Acknowledgments**

This work was supported by NYUAD Research Enhancement Fund and UAE department of education and knowledge, ADEK Awards for Research Excellence (AARE 2018). The authors are thankful to NYUAD Photonics and Core Technology Platform Facility (CTP) for the analytical, material characterization, devices fabrication and testing.

**Disclosures**

The authors declare no conflicts of interest.

**Supporting Information**

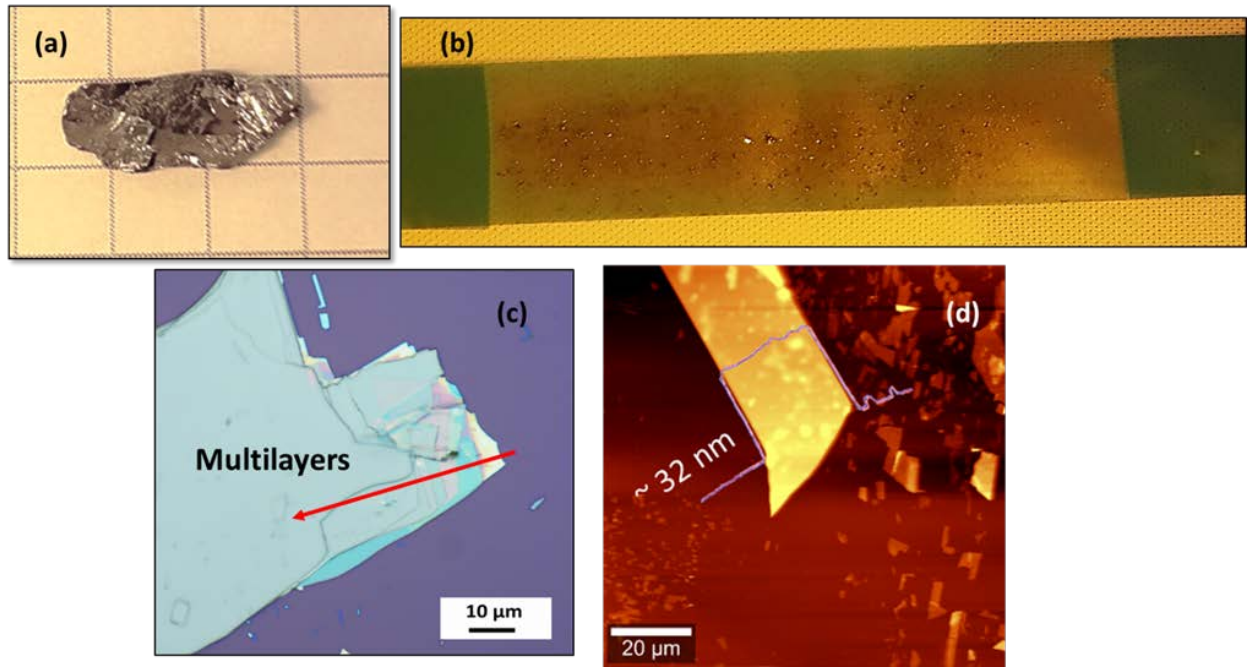

Figure S1. 2D GeAs exfoliation from bulk crystals (a) bulk stone commercially available from 2D semiconductors (b) exfoliated flakes on the type (c) optical image showing multilayered 2D GeAs (d) AFM scan showing ~ 32 nm flake thickness.

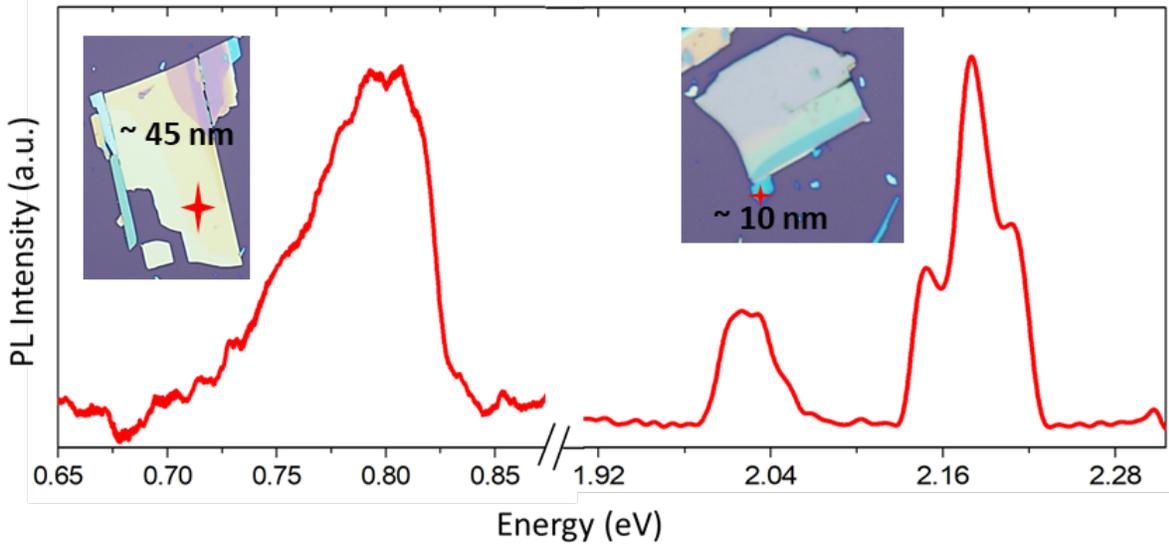

Figure S2. PL spectra of 2D GeAs at different thickness, the red stars indicate the place where the laser is focused. Thick flakes showed NIR peak while thin flakes have emission in visible.

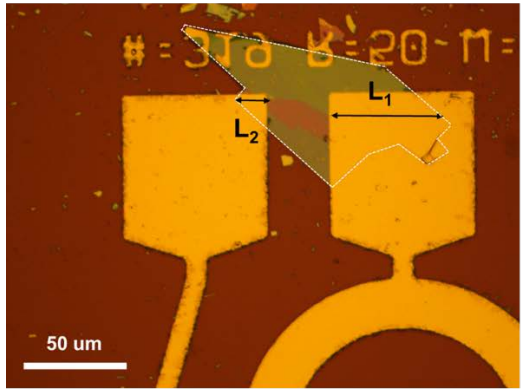 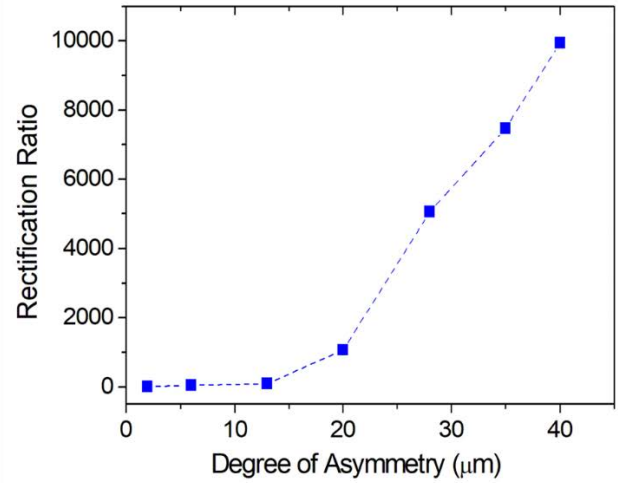

Figure S3. Electrical characteristic of GeAs Schottky photodiode (a) Optical image showing the degree of asymmetry of the contact area (ΔL=$L_1$-$L_2$) (b) rectification ratio as a function of ΔL.

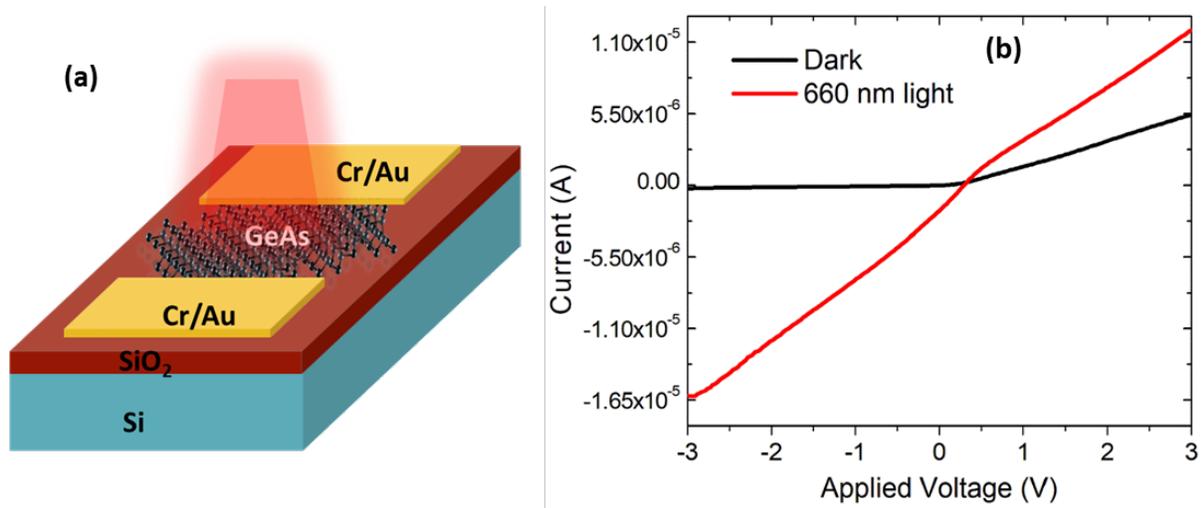

Figure S4. Photo-induced characteristic of GeAs Schottky photodiode (a) schematic of the device structure (b) I-V curves on a linear scale under 660 nm laser illumination along with their dark state, inset showing the asymmetric geometry of the flake contact area ( red is the developed resist before metal deposition step).